# Focusing large spectral bandwidths through scattering media

Arturo G. Vesga[1,&], Matthias Hofer[1], Naveen Kumar Balla[1,#], Hilton B. De Aguiar[2], Marc Guillon[3], and Sophie Brasselet[1,*]

[1]Aix Marseille Univ, CNRS, Centrale Marseille, Institut Fresnel, F-13013 Marseille, France
[2]Département de Physique, Ecole Normale Supérieure/PSL Research University, CNRS, 24 rue Lhomond, 75005 Paris, France [3]Currently with the Department of Electronic Journals
[3]Paris Descartes University, 45, rue des Saints Pères, 75270 Paris Cedex 06, France
[&] Present address: Biofisika Institute (CSIC, UPV/EHU), University of the Basque Country (UPV/EHU), P.O. Box 644, 48080 Bilbao, Spain
[#] Present address: CRANN, School of Physics, Trinity College Dublin, The University of Dublin, College Green, Dublin 2, Ireland
*sophie.brasselet@fresnel.fr

**Abstract:** Wavefront shaping is a powerful method to refocus light through a scattering medium. Its application to large spectral bandwidths or multiple wavelengths refocusing for nonlinear bio-imaging in-depth is however limited by spectral decorrelations. In this work, we demonstrate ways to access a large spectral memory of a refocus in thin scattering media and thick forward-scattering biological tissues. First, we show that the accessible spectral bandwidth through a scattering medium involves an axial spatio-spectral coupling, which can be minimized when working in a confocal geometry. Second, we show that this bandwidth can be further enlarged when working in a broadband excitation regime. These results open important prospects for multispectral nonlinear imaging through scattering media.

## 1. Introduction

Nonlinear label-free bio-imaging is strongly developing as a promising tool for in situ biomedical applications. Reaching large penetration depths that surpass several hundreds of micrometers in tissues is however still a major challenge [1][2]. Adaptive optics tools are able to generate nonlinear signals at shallow depths only, since they primarily make use of ballistic photons that rapidly vanish when reaching several scattering mean free paths [3][4][5]. When highly scattering regimes dominate, other imaging strategies are required. Wavefront shaping is an interesting approach that makes use of all scattered paths to modify the initial random intensity speckle pattern into a refocus, using optimization [6] or measurement of the transmission matrix of the medium [7]. Initially developed for imaging through scattering media, this method is potentially applicable in reflection [8] which is of high importance for biological imaging. Recently, it has been used to generate efficient second harmonic generation (SHG) signals from narrow band pulses through anisotropic biological scattering media [9].

Many nonlinear optical interactions used for tissue imaging are however more complex than SHG since they involve polychromatic configurations with more than one beam, sometimes in a broadband ultra-short pulsed regime. The most frequent examples are multicolour multiphoton fluorescence imaging [10][11], four wave mixing, coherent anti stokes Raman scattering (CARS) [12], and sum frequency generation (SFG) [13]. These situations are very challenging for wavefront shaping since coherent manipulation of waves through a scattering medium is only applicable within a wavelength range below the spectral bandwidth of the medium [14][[15][16][17]. This bandwidth, which represents the number of independent spectral modes of the medium, is generally assessed by its speckle spectral correlation bandwidth [16][18] and does not surpass a few tens of nanometers in millimetre-thick biological media [19].

As a consequence, the spectral memory of the refocus formed by wavefront shaping, e.g. its robustness to spectral detuning, is limited by the medium spectral bandwidth. In this work, we show that two additional parameters strongly influence the spectral memory of a refocus in a non-trivial way. First, the axial spatio-spectral coupling of the waves propagation in thin or forward- scattering media can strongly reduce the accessible refocus spectral bandwidth. We evaluate the role of the axial position of the refocus plane along the propagation direction on the refocus spectral bandwidth and show that when the imaging system is in the confocal geometry (e.g. the focal points of the excitation and collection optics in free space coincide), the largest spectral bandwidths can be obtained with minimal spatio-spectral decorrelations. Second, we investigate the effect of the spectral bandwidth of the incident laser beam and demonstrate that broadband refocussing is able to enlarge the accessible refocus spectral bandwidth beyond the spectral bandwidth of the medium, as a direct consequence of the coherent regime of interferences occurring when the refocus is formed.

## 2. 3D speckle spectral correlations

The monochromatic illumination of a scattering medium leads to the formation of a speckle, which results from interferences of spatially random nature, between the different paths undergone by light within the medium. Under monochromatic illumination with a tuneable incident wavelength, the spectral bandwidth of the refocus obtained by wavefront shaping is expected to be the same as the speckle spectral bandwidth [14]. However since in thin scattering media the spectral properties of a speckle are also strongly correlated to its spatial axial expansion [20], the axial dimension can have strong consequences on the measured speckle and refocus spectral bandwidths. The importance of this effect is studied in what follows.

The experimental set up depicted in Fig. 1a permits to measure speckle as well as refocussing spectral decorrelations in a microscopy imaging geometry. The incident beam is produced by a tunable OPO (Coherent) pumped by a picosecond laser (pump 1031nm, 3ps/pulse, 0.1 nm spectral width, APE) expanded with a telescope. This source can be considered as close to monochromatic even for thin scattering media. Light is sent on a 256x256 pixel reflective spatial light modulator (SLM, Boulder Nonlinear Systems), conjugated to the back focal plane of the excitation objective O1 (Olympus LCPLN-IR 20x, 0.45 NA)  (Fig. 1a). Note that the SLM can be considered as achromatic in the spectral ranges used in the experiment. The beam is focused on the sample plane, which is imaged by the objective O2 (Olympus PlanFLN 40x, 0.6 NA) on a CMOS camera (Blackfly U3-23S6M-C, FLIR). The sample is placed between the objectives with a fixed holder. O2 can be translated along the optical axis using a translation stage to explore the axial properties of both speckles and refocussed points. For refocussing behind the scattering medium, the transmission matrix (TM) of the medium is measured as described in [7]. The SLM is used to phase-tune part of the incident wave front, the other part serving as a reference beam. Self-reference interferometry using a wave front decomposition on Hadamard bases is used to extract the TM matrix relation between the incident and out-going field modes [7]. We use 1024 (32x32 macropixels on the SLM) Hadamard bases to scan the TM. Note that this TM also contains possible contributions from the illumination and detection optics. The long focal length of objective O1 (8.3 mm) allows to place the scattering medium at a chosen position from its focal plane F within millimetre distances (Fig. 1b). In addition the refocus plane R, where the TM is measured, is not necessarily the focal plane F (Fig. 1b).

Speckles measured at a few mm from a polycarbonate diffuser (Newport, 10°, Light Shaping Diffuser) are displayed in Fig. 1c for different wavelengths separated by 6 nm, starting from

$\lambda_0 = 828$ nm. This wavelength shift induces visible decorrelations between the obtained speckles. We evaluated how this speckle decorrelation affects a refocussing experiment by wavefront shaping. After the TM of the scattering medium is measured at $\lambda_0$ to produce a refocus in the reference plane R, the wavelength is tuned. At $\lambda_1 = 822$ nm and in a more pronounced way at $\lambda_2 = 816$ nm, the refocus is seen to quickly degrade (Fig. 1d). This is expected from the observed speckle spectral decorrelation (Fig. 1c), which indicates a similar degree of decorrelation in the measured TM [14]. Importantly however, moving the O2 objective away from the R plane along the axial propagation direction allows recovering a better refocus quality, even though the enhancement performances are slightly degraded (Fig. 1e). Although chromatic aberrations of lenses contribute to this effect, they cannot entirely explain the large axial shifts observed. Indeed the observed axial shift is 22 μm for a wavelength shift of 12 nm, while pure chromatic effects, measured in a similar geometry without the scattering medium, lead to a spatial displacement of less than 6 μm. As a reference for chromatic aberrations in the system, an axial shift dependence of 0.45 μm/nm is measured when focussing light at the same R plane without the scattering medium.

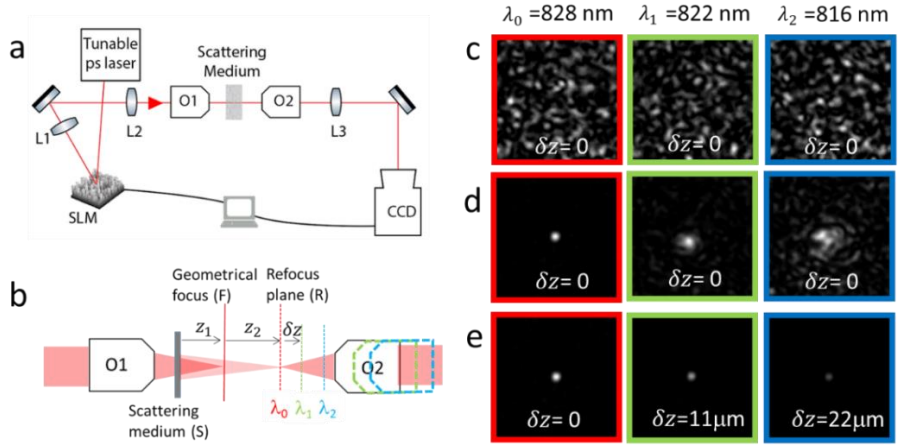

Fig. 1. (a) Experimental layout. O1, O2: objectives. SLM: spatial light modulator. L1,L2 (lenses) are used to image the SLM on the back focal plane of O1. L3 (lens) is used to image the refocus plane on the CCD camera. (b) Positioning of the sample and objectives. $z_1$ = SF is the distance between the scattering medium (here a diffuser) and the geometrical focus (F). $z_2$ = FR is the distance between F and the refocus plane R at $\lambda_0$. $\delta z$ is the distance between R and the new position of the refocus plane when the wavelength $\lambda$ is shifted from its initial value $\lambda_0$. (c) Speckle images obtained at the fixed plane R for several wavelengths separated by 6 nm, for $z_1$ = 2 mm and $z_2$ = 1 mm. (d) Wavefront shaping experiment producing a refocus in R. Once the TM is measured for $\lambda_0$, the incident wavelength is tuned to $\lambda_1$ and $\lambda_2$, producing distorted focus images at $\delta z = 0$. (e) Translating the image planes by $\delta z$ allows to recover a focus at $\lambda_1$ and $\lambda_2$.

We verified that the recovery of the focus quality when shifting the image plane by $\delta z$ is also accompanied by an increase of speckle correlations. Figure 2 represents the speckle spectral decorrelation curve measured at the reference refocus plane R, obtained by tuning $\lambda$ away from the reference wavelength $\lambda_0 = 828$ nm. In this experiment, the scattering medium is placed a few mm's before the geometrical focus F and the plane R is close to F (Fig. 2a). Note that a limited amount of spectral points are measured per curve to guaranty the stability of the set-up during wavefront shaping and spectral/axial tuning. In the plane R, a speckle decorrelation bandwidth $\Delta\lambda$ of about 18 nm is obtained (Fig. 2b, red curve). When varying $\lambda$, an axial shift of the image plane away from the reference plane R permits however to recover a higher degree of correlation with the reference speckle (Fig. 2b, blue curve). We define as '3D bandwidth' $\Delta\lambda_{3D}$ the speckle spectral width obtained when correlation measurements are performed at the planes with highest correlation for each tuned incident wavelength. $\Delta\lambda_{3D}$ is about 87 nm in the configuration described in Fig. 2 (Fig. 2b, blue curve), almost a factor 5 above the initial 2D bandwidth. Interestingly, the 3D spectral decorrelations curves of both speckle and refocus (Fig. 2b, green curve), are relatively similar, and it is found that the highest correlation plane for the speckle is also the plane for recovering an optimized focus when the wavelength is tuned away from $\lambda_0$. The axial translation extent necessary to cover the full 3D spectral width of an optimized focus is of the order of tens of µm in the present configuration (Fig. 2c).

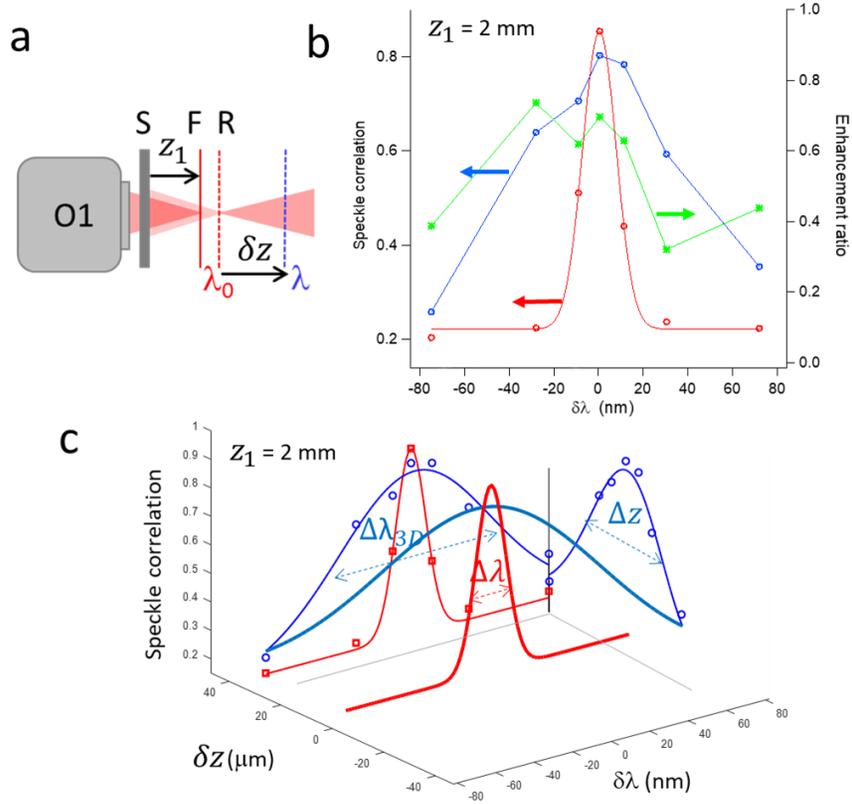

Fig 2. Measured speckle correlations in 2D, 3D, and enhancement ratio at the best focus plane, in a diffuser. (a) The reference R plane position is set close to the geometrical focus plane ($z_2 \sim$ 120 µm). The scattering medium output surface is positioned at $z_1 = 2$ mm from the geometrical focus F. (b) Measured speckle correlations obtained when tuning the incident wavelength away from $\lambda_0 = 828$ nm: in 2D (red markers), 3D (blue markers) and enhancement ratio at the best focus plane (green markers). The wavelength axis represents the shift $\delta\lambda$ of the measured wavelength with respect to the reference wavelength $\lambda_0$. The continuous line is a fit of the 2D experimental data by a Gaussian function. Here, the refocus 2D bandwidth is not represented due to the fast degradation of the refocus quality when changing the incident wavelength, making the estimation of the enhancement ratio imprecise. (c) Same configuration represented in a 3D plot where the translation distance $\delta z$, necessary to find the best focus at wavelengths $\lambda$, is represented as an additional axis. The speckle correlation 2D (red) and 3D (blue) experimental points (markers) and fits (continuous lines) are represented both in the $\lambda$-space and in the $z$-space, as well as in 3D (continuous thick lines crossing the graph). In this situation the measured spectral decorrelation bandwidths are respectively $\Delta\lambda = 18$ nm in 2D and $\Delta\lambda_{3D} = 87$ nm in 3D, while the spatial extent of the 3D bandwidth is $\Delta z = 30.5$ µm. Note that this size is larger than the speckle axial grain size (which is less than 8 µm).

Overall, this experiment shows that the spectral bandwidth of the diffuser cannot be solely defined and characterized at a single plane R. It is indeed affected by the presence of spatio-spectral correlations in the axial direction $z$, which manifest themselves in the speckle formed after the propagation in the medium, and thus in its refocussing capabilities.

Note that the chromatic nature of spatial speckle correlations can take complex forms, since axial and lateral dimensions contribute differently to speckle propagation. In addition to axial correlations which are the center of this work, a lateral dilatation of the speckle is also expected when $\lambda$ and $z$ are tuned, which originates from the convergent illumination of the scattering medium [21][22]. This lateral dilatation of the speckle may explain the slight discrepancy between the blue and green curves plotted in Fig. 2b. Nevertheless, lateral effects along the $(x, y)$ transverse dimensions are non-negligible only away from the propagation axis, as previously observed in nonlinear refocus-scanning imaging [23].

### 3. Spatial extent of the refocus 3D bandwidth

To get further insight in the spatio-spectral characteristics of the refocus along the axial direction, we quantified the spatial extent along $z$ of the 3D refocus spectral bandwidth, for different imaging geometries. Figure 3 shows how this spatial extent varies with the distance $z_2$, for a scattering medium placed at a few mm from the geometrical focus F ($z_1 = 5$ mm) (Fig. 3a). When $z_2$ is a few millimeters, the full spatial extent $\Delta z$ of the refocused point (necessary to scan the full 3D spectral bandwidth $\Delta\lambda_{3D}$) reaches hundreds of µm (Fig. 3b). This spatio-spectral dependence is symmetric with respect to the reference plane position, which shows a singular behaviour at this specific point (Fig. 3c). Removing the scattering medium in similar convergent illumination shows that less than of 30% of the measured displacement values is attributed to chromatic aberrations. Note that whatever the axial position of the produced refocus, its 3D spectral extent is limited by the intrinsic spectral bandwidth of the diffuser, which is measured to be around 87 nm from Fig. 2.

Figure 3 shows that increasing the distance $z_2$ (and similarly $z_1$) leads to an axial expansion of the spectral bandwidth. In contrast when the scattering medium and observation planes are at the geometric focus position ($z_1 = z_2 = 0$), e.g. when the system is set in the confocal geometry for both excitation and detection objectives, the spatial extent of the 3D spectral bandwidth is minimized. Note that removing the excitation objective to produce a planar wave illumination leads to a 3D bandwidth that cannot be confined in one given plane as expected from free space speckle propagation.

The dependence observed in Fig. 3 is attributed to the presence of spatio-spectral coupling along the axial direction of the wave propagation after the scattering medium. This coupling naturally takes the form of phase contributions in the wave propagation taking place at the exit of the medium, which contains in particular a parabolic form when a thin medium is illuminated under convergence illumination [21]. We model below the simple case of scattering by a thin diffuser to emphasize the role of spatio-spectral coupling. Under coherent convergent illumination, the formation of a speckle is the result of diffraction which takes a form that depends both on $\lambda$ and axial propagation distance $z$ in an intrinsically coupled manner [20]. In the Fresnel regime, the observed intensity follows the Fresnel-Kirchhoff equation for a convergent illumination [21]:

$$I(u,v) = \left| \iint G(x,y) \exp\left(\frac{-i\pi}{\lambda z}\frac{(z-z_1)}{z_1}(x^2+y^2)\right) \exp\left(\frac{-i2\pi}{\lambda z}(ux+vy)\right) dxdy \right|^2 \qquad (1)$$

With $I(u,v)$ the intensity measured in the image plane of coordinates $(u,v)$, placed at a distance $z$ from the scattering thin diffuser made of a phase-only transmission mask $G(x,y)$. $z = (\delta z + z_1 + z_2)$ is the distance between the scattering medium and the observation plane for the wavelength $\lambda$ (Fig. 3a).

In Eq. (1), simple spatio-spectral correlations properties can be found in two extreme cases. First, under planar wave illumination, the term in $z/z_1$ in the first complex exponential in Eq. (1) vanishes and the remaining term is invariant with respect to the dependence $\lambda z = const.$ [21] (where $const.$ is a constant number). This situation has permitted depth access in speckle imaging using spatio-spectral correlations [24][25]. Second, under convergent illumination and observation close to the optical axis ($(u,v) \approx (0,0)$), the second term vanishes and a parabolic phase remains, leaving an invariance with respect to the dependence $\lambda z.z_1/(z-z_1) = const.$ In the context of this work we follow this second situation, the refocus being measured on the optical axis. Note that far from the optical axis, this equation emphasizes extra spatio-spectral dilatations in the lateral dimension $(x,y)$ as mentioned above [22][21].

When $\lambda$ or $z$ are varied, the case $(u,v) \approx (0,0)$ permits to evidence planes of constant spatio-spectral correlations which follow $\lambda z.z_1/(z-z_1) = const.$ This dependence is displayed in Fig. 3c together with the experimental data, fixing $(z_1, z_2)$ as the distances used in the experiment. The interpretation of our results with the Fresnel propagation model provides a good qualitative agreement with the experimentally observed spatio-spectral coupling in thin scattering medium. In particular, it displays a symmetry with respect to $\delta z = 0$, and a minimal spatial extent of the spectral variables when $z_2 = 0$, as observed experimentally. Note however that we assumed in the model that the transmission coefficient of the diffuser is achromatic, which is an approximation.

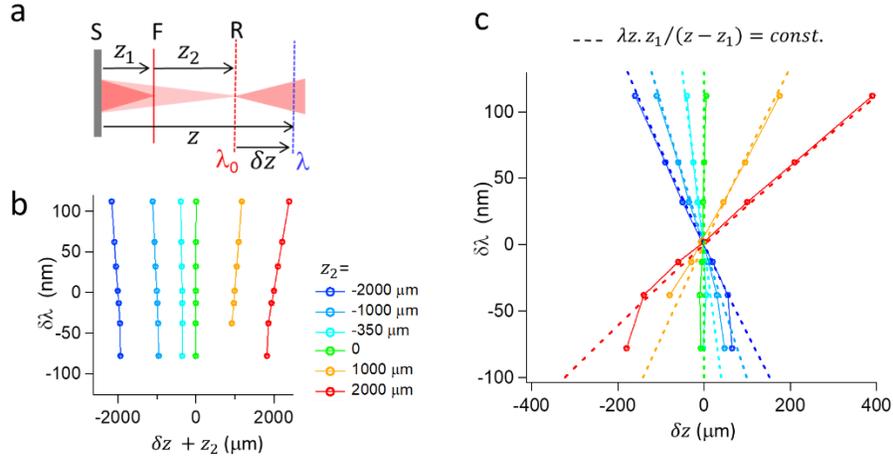

Fig 3. Relation between wavelength detuning $\delta\lambda$ and axial distance between foci. (a) Experimental configuration showing the different planes involved, $\delta z$ is the translation required to recover an optimized focus at $\lambda$ (ranging from 750 nm to 940 nm), relative to the reference plane R of refocus at $\lambda_0 = 828$ nm. The distance from the scattering medium to the geometrical focus is $z_1 = 5$ mm, and the refocus plane position R is varied between $z_2 = -2$ mm and 2 mm. (b) Displacement relative to the plane R ($\delta z + z_2$) required for wavelengths shifts $\delta\lambda$ defined by ($\lambda = \lambda_0 + \delta\lambda$), for different values $z_2$. (c) Same data shown as a function of $\delta z$. The dashed lines represent a function that follows the law $\lambda z.z_1/(z-z_1) = const.$

The results of Fig. 3 lead to two important conclusions. First, despite the presence of the diffuser, the light beam keeps the 'memory' of the geometrical focus of the objective lens

(manifested as a parabolic phase) when the wavelength is shifted. This is expected by the fact that the scattering regime explored here is far from a multiple scattering situation, which would randomly scramble propagation phase contributions. Second, in a thin scattering medium the spectral memory of the refocus formed by wavefront shaping at a given plane, depends on the axial position of this plane. It is in particular maximized when this planes coincides with the geometrical focus of the excitation optics, e.g. when the set-up is set in a confocal geometry for both excitation and detection objectives. At last, this result shows that the characterization of the spectral bandwidth of a thin scattering medium may be underestimated if only measured in 2D away from the confocal geometry.

## 4. Spatial extent of the refocus 3D bandwidth in thin and anisotropic scattering media

The previous measurements were performed using a diffuser as a scattering medium, which is close to a thin scattering medium, manifested also by a large angular memory effect [26][27]. We evaluated in Fig. 4 the 2D and 3D spectral bandwidths as well as the axial expansion of the 3D bandwidth in different media, either thin or exhibiting an anisotropic scattering nature. In such cases, we expect the memory of the excitation geometrical focus to be present to some degree, which should preserve the spatio-spectral coupling effect observed above. First, a fixed 1mm thick opaque brain slice from a mouse (coronal cross section) is used as biological sample. This sample exhibits 2D and 3D spectral bandwidths that are similar to those of a diffuser. The enhancement of the spectral bandwidth when moving from a 2D measurement to 3D is about 5 to 10 fold. The scattering mean free path of a biological tissue is about 100 μm, but its anisotropic factor *g* close to 1 makes its transport mean free path larger than 1 mm [28], which is the thickness explored here. We compared these values to scattering media exhibiting a lower anisotropy factor, but thin with respect to their transport mean free path. These samples are made of diluted $TiO_2$ nanoparticles of 500nm size deposited on a coverslip with thicknesses 20 μm and 40 μm, embedded in a glue of refraction index close to 1.5. This permits to reduce their scattering power, and in particular to generate measurable 2D bandwidths, above the 0.1 nm spectral width of the laser used. Even though this regime is thus far from being diffusive, both the 2D and 3D bandwidths measured in the $TiO_2$ samples are lower than in the biological tissue, which is attributed to their lower thickness-to-transport mean free path ratio. As a control a very thick $TiO_2$ sample is also reported, where spectral bandwidths are measured to be below 2 nm. In this case, no spatio-spectral coupling effect was measurable, as expected from a regime where the geometrical phase memory of the excitation wavefront is lost.

A common slope $\Delta\lambda_{3D}/\Delta z$ can be drawn out of the measurements performed in the thin and anisotropic media (Fig. 4), which is of the order of about 100nm/30μm for a given $z_1$ distance. The axial expansion of the spectral bandwidth does not reflect the medium but rather propagation properties of the speckle generated by the medium. In particular the measured slope is expected to be related to $z_1$ (Fig. 3). Figure 4 also shows that in anisotropic scattering media, very large 3D bandwidth can be obtained. This increased capacity for spectral memory can be related to the previously reported large geometrical memory effect in such media [29], which has been generalized to a shift/tilt memory effect [30] combining angular [31] and spatial [32] memory effects.

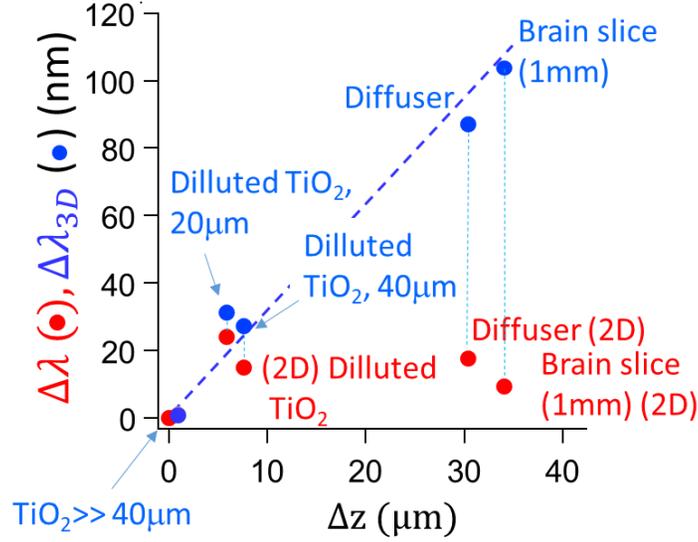

Fig 4. 2D (red) and 3D (blue) spectral bandwidths reported as a function of the spatial extent $\Delta z$ of $\Delta\lambda_{3D}$ for different media (see text) in the geometrical configuration $z_1 = 5$ mm. The refocus (reference) R plane position is set close to the scattering medium surface ($z_2 \sim 120$ μm). The blue dashed line is added as a guide line.

## 5. Enlarging the 3D refocus bandwidth using broadband wavefront shaping

All experiments above were performed close to a monochromatic regime. We explore in what follows the consequence of use of a broadband excitation pulse, which is generally required for nonlinear optical generation. A short pulse is expected to be strongly distorted at the exit of a scattering media, as a consequence of the randomization of propagation lengths in the medium. To control short pulsed laser light with large bandwidths in scattering media, several strategies have been developed, based on the direct optimization of a nonlinear signal [33][34], the control of the incident pulse in the time domain [35][36], or on the independent manipulation of different spectral channels [37]. A faster and more direct approach has recently been achieved by measuring the transmission matrix of the medium in the broadband regime [38][39]. The measured 'broadband TM' has to be distinguished from the 'multispectral TM' reconstructed from a collection of individual monochromatic TM's. The broadband TM has been shown in particular to preserve the short pulse nature of the refocus, due to an intrinsic coherent time gating effect produced by self-reference interferometry [38], provided that the medium bandwidth is comparable with the laser bandwidth. As a consequence a broadband refocus exploits favourably short propagation paths in a scattering medium [40][38]. This effect, which has been exploited for nonlinear imaging [9] and polarization-preserved imaging [40], should also have consequences on the spectral bandwidth of the refocus.

To assess the spectral memory of such a broadband refocus, we use the confocal geometry mentioned above ($z_2 = z_1 = 0$). and measure the TM of the medium under short pulsed, broadband conditions. The laser is now a tunable OPO (Coherent) pumped by a Ti:Sa oscillator (150 fs/pulse, 80 MHz) (Coherent) with a spectral width of about 8 nm. Experiments were first performed at various $z_1$ distances, the reference plane being at the geometrical focus of the objective O1 (Fig. 5a). The measured 3D bandwidth is considerably enhanced (Fig. 5b) and reaches $\Delta\lambda_{3D}$ values of 260 nm, about a factor 3 above the bandwidth obtained in the ps (close to monochromatic) regime. This increase is attributed to the coherent gating effect produced by the measurement of the broadband TM transmission matrix [38], which intrinsically enlarges spectral capabilities as recently demonstrated by time gating [41]. When tuning the $z_2$ distance ($z_1$ is fixed at 2 mm from the diffuser) (Fig. 5c), the obtained 3D bandwidth is very similar for all $z_2$ distances (Fig. 5d), confirming that the obtained 3D bandwidth value is an intrinsic property of the medium. Note that this spectral bandwidth is twice as large as the speckle correlation bandwidth (about 80 nm in the confocal geometry, Fig. 5e), which keeps a similar magnitude as the one measured in the close-to-monochromatic ps regime. This is due to the fact that while a refocus is constructed in coherent conditions, the speckle measurement is performed on intensities. At last, the spatial dependence of the spectrally large refocus (Fig. 5f) shows that the axial expansion of $\Delta\lambda_{3D}$ is minimized around F when the reference plane is at the geometrical focus, while it extends hundreds of μm away when $z_2$ increases. This behaviour is similar to what was observed in the ps, close-to-monochromatic regime (Fig 5f).

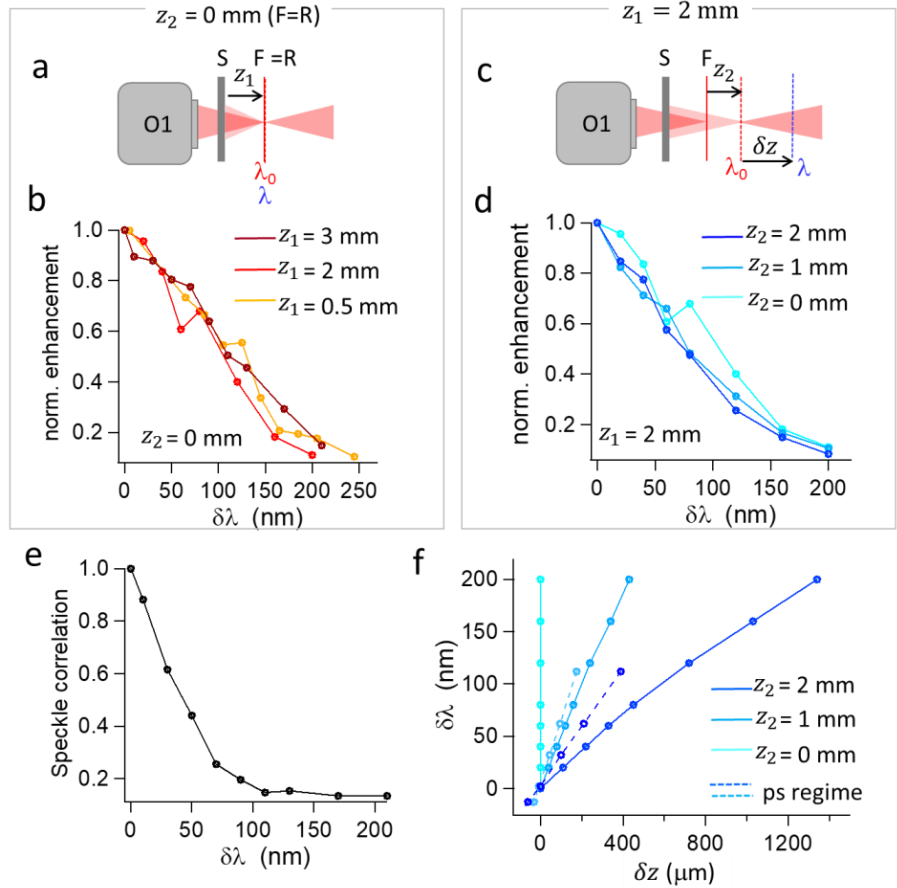

Fig 5. Broadband refocussing in the multispectral regime, using a 150 fs source centered on $\lambda_0$ = 785 nm, and a diffuser as the scattering medium. (a) Scheme for evaluating the dependence of $\Delta\lambda_{3D}$ as a function of the $z_1$ (= SF) distance. The reference refocus plane R for $\lambda_0$ is at the geometrical focus position. (b) 3D spectral correlation of refocussed beams, for different SF distances. (c) Scheme for evaluating the dependence of $\Delta\lambda_{3D}$ as a function of the FR distance. (d) $\Delta\lambda_{3D}$ obtained for different FR distances ($z_1$ = 2 mm). (e) 2D speckle correlation bandwidth obtained in the confocal geometry (this bandwidth is similar to 3D speckle correlation bandwidths when the system is not in the confocal geometry). (f) Relation between spectral detuning and the axial displacement of the best refocus plane, for different $z_2$ (= FR) distances. The data are compared to the values obtained in Fig. 3.

## 6. Conclusion

We have shown that spatio-spectral coupling in thin and anisotropic scattering media has important consequences on the spectral memory of a refocus formed by wavefront shaping. The axial expansion of refocus spectral correlations is in particular minimized when working in the confocal excitation/detection geometry, leading to large spectral bandwidth limited only by the medium. This work demonstrates moreover that under broadband illumination conditions, it is possible to refocus wavelengths that are distant from each other by a large amount. This situation is typically met in four wave mixing and coherent anti Stokes Raman microscopy, or in sum frequency nonlinear excitations. In non-confocal situations, it should be possible to use a single TM for the refocussing of shifted wavelengths, making the medium characterization for further imaging much faster than in sequential-optimization based methods. That could be done by compensating for the axial shift induced by speckle correlation properties using additional correction wavefronts.

## 7. Funding, acknowledgments, and disclosures

*7.1 Funding*

This work has been supported by ANR-15-CE19-0018-01 (MyDeepCARS), ANR-10-INBS-04-01 (France-BioImaging), and the A*Midex Interdisciplinary project *Neurophotonics*. H.B.A. was supported by LabEX ENS-ICFP: ANR-10-LABX-0010/ANR-10-IDEX-0001-02 PSL*.

*7.1 Acknowledgments*

We thank Christian Soeller (Exeter University, UK) for providing the fixed mouse brain slice samples.

## References

1. K. Wang, W. Sun, C. T. Richie, B. K. Harvey, E. Betzig, and N. Ji, "Direct wavefront sensing for high-resolution in vivo imaging in scattering tissue," Nat. Commun. **6**(1), 7276 (2015).

2. J.-H. Park, W. Sun, and M. Cui, "High-resolution in vivo imaging of mouse brain through the intact skull.," Proc. Natl. Acad. Sci. U. S. A. **112**(30), 9236–41 (2015).

3. R. K. Tyson, *Principles of Adaptive Optics* (n.d.).

4. M. J. Booth, "Adaptive optics in microscopy.," Philos. Trans. A. Math. Phys. Eng. Sci. **365**(1861), 2829–43 (2007).

5. J. A. Kubby, *Adaptive Optics for Biological Imaging* (CRC Press, 2013).

6. I. M. Vellekoop and A. P. Mosk, "Focusing coherent light through opaque strongly scattering media," Opt. Lett. **32**(16), 2309 (2007).

7. S. M. Popoff, G. Lerosey, R. Carminati, M. Fink, A. C. Boccara, and S. Gigan, "Measuring the Transmission Matrix in Optics: An Approach to the Study and Control of Light Propagation in Disordered Media," Phys. Rev. Lett. **104**(10), 100601

(2010).

8. A. Badon, D. Li, G. Lerosey, A. C. Boccara, M. Fink, and A. Aubry, "Smart optical coherence tomography for ultra-deep imaging through highly scattering media," Sci. Adv. **2**(11), e1600370 (2016).

9. H. B. De Aguiar, S. Gigan, and S. Brasselet, "Enhanced nonlinear imaging through scattering media using transmission-matrix-based wave-front shaping," Phys. Rev. A - At. Mol. Opt. Phys. **94**(4), (2016).

10. C. Stringari, L. Abdeladim, G. Malkinson, P. Mahou, X. Solinas, I. Lamarre, S. Brizion, J.-B. Galey, W. Supatto, R. Legouis, A.-M. Pena, and E. Beaurepaire, "Multicolor two-photon imaging of endogenous fluorophores in living tissues by wavelength mixing," Sci. Rep. **7**(1), 3792 (2017).

11. J. W. Lichtman, J. Livet, and J. R. Sanes, "A technicolour approach to the connectome," Nat. Rev. Neurosci. **9**(6), 417–422 (2008).

12. J.-X. Cheng and X. S. Xie, "Vibrational spectroscopic imaging of living systems: An emerging platform for biology and medicine.," Science **350**(6264), aaa8870 (2015).

13. A. Hanninen, M. W. Shu, and E. O. Potma, "Hyperspectral imaging with laser-scanning sum-frequency generation microscopy," Biomed. Opt. Express **8**(9), 4230 (2017).

14. F. van Beijnum, E. G. van Putten, A. Lagendijk, and A. P. Mosk, "Frequency bandwidth of light focused through turbid media," Opt. Lett. **36**(3), 373 (2011).

15. E. A. Shapiro, T. M. Drane, and V. Milner, "Prospects of coherent control in turbid media: Bounds on focusing broadband laser pulses," Phys. Rev. A **84**(5), 053807 (2011).

16. N. Curry, P. Bondareff, M. Leclercq, N. F. van Hulst, R. Sapienza, S. Gigan, and S. Grésillon, "Direct determination of diffusion properties of random media from speckle contrast," Opt. Lett. **36**(17), 3332 (2011).

17. H. P. Paudel, C. Stockbridge, J. Mertz, and T. Bifano, "Focusing polychromatic light through strongly scattering media," Opt. Express **21**(14), 17299 (2013).

18. D. Andreoli, G. Volpe, S. Popoff, O. Katz, S. Grésillon, and S. Gigan, "Deterministic control of broadband light through a multiply scattering medium via the multispectral transmission matrix," Sci. Rep. **5**(1), 10347 (2015).

19. O. Katz, P. Heidmann, M. Fink, and S. Gigan, "Non-invasive single-shot imaging through scattering layers and around corners via speckle correlations," Nat. Photonics **8**(10), 784–790 (2014).

20. J. W. Goodman, *Speckle Phenomena in Optics : Theory and Applications* (Roberts & Co, 2007).

21. M. May, "Information inferred from the observation of speckles," J. Phys. E. **10**(9), 849–864 (1977).

22. E. Small, O. Katz, Y. Eshel, Y. Silberberg, and D. Oron, "Spatio-temporal X-wave," Opt. Express **17**(21), 18659 (2009).

23. M. Hofer and S. Brasselet, "Manipulating the transmission matrix of scattering media for nonlinear imaging beyond the memory effect," Opt. Lett. **44**(9), 2137 (2019).

24. X. Xu, X. Xie, A. Thendiyammal, H. Zhuang, J. Xie, Y. Liu, J. Zhou, and A. P.

Mosk, "Imaging of objects through a thin scattering layer using a spectrally and spatially separated reference," Opt. Express **26**(12), 15073 (2018).

25. J. Liang, J. Cai, J. Xie, X. Xie, J. Zhou, and X. Yu, "Depth-resolved and auto-focus imaging through scattering layer with wavelength compensation," J. Opt. Soc. Am. A **36**(6), 944 (2019).
26. O. Katz, E. Small, Y. Guan, and Y. Silberberg, "Noninvasive nonlinear focusing and imaging through strongly scattering turbid layers," Optica **1**(3), 170 (2014).
27. M. Hofer, C. Soeller, S. Brasselet, and J. Bertolotti, "Wide field fluorescence epi-microscopy behind a scattering medium enabled by speckle correlations," Opt. Express **26**(8), 9866 (2018).
28. S. L. Jacques, "Optical properties of biological tissues: a review," Phys. Med. Biol. **58**(11), R37–R61 (2013).
29. S. Schott, J. Bertolotti, J.-F. Léger, L. Bourdieu, and S. Gigan, "Characterization of the angular memory effect of scattered light in biological tissues," Opt. Express **23**(10), 13505 (2015).
30. G. Osnabrugge, R. Horstmeyer, I. N. Papadopoulos, B. Judkewitz, and I. M. Vellekoop, "Generalized optical memory effect," Optica **4**(8), 886 (2017).
31. I. Freund, M. Rosenbluh, and S. Feng, "Memory Effects in Propagation of Optical Waves through Disordered Media," Phys. Rev. Lett. **61**(20), 2328–2331 (1988).
32. B. Judkewitz, R. Horstmeyer, I. M. Vellekoop, I. N. Papadopoulos, and C. Yang, "Translation correlations in anisotropically scattering media," Nat. Phys. **11**(8), 684–689 (2015).
33. O. Katz, E. Small, Y. Bromberg, and Y. Silberberg, "Focusing and compression of ultrashort pulses through scattering media," Nat. Photonics **5**(6), 372–377 (2011).
34. J. Aulbach, B. Gjonaj, P. Johnson, and A. Lagendijk, "Spatiotemporal focusing in opaque scattering media by wave front shaping with nonlinear feedback," Opt. Express **20**(28), 29237 (2012).
35. D. J. McCabe, A. Tajalli, D. R. Austin, P. Bondareff, I. A. Walmsley, S. Gigan, and B. Chatel, "Spatio-temporal focusing of an ultrafast pulse through a multiply scattering medium," Nat. Commun. **2**(1), 447 (2011).
36. M. Mounaix, H. Defienne, and S. Gigan, "Deterministic light focusing in space and time through multiple scattering media with a time-resolved transmission matrix approach," Phys. Rev. A **94**(4), 041802 (2016).
37. M. Mounaix, D. Andreoli, H. Defienne, G. Volpe, O. Katz, S. Grésillon, and S. Gigan, "Spatiotemporal Coherent Control of Light through a Multiple Scattering Medium with the Multispectral Transmission Matrix," Phys. Rev. Lett. **116**(25), 253901 (2016).
38. M. Mounaix, H. B. de Aguiar, and S. Gigan, "Temporal recompression through a scattering medium via a broadband transmission matrix," Optica **4**(10), 1289 (2017).
39. M. Mounaix, D. M. Ta, and S. Gigan, "Transmission matrix approaches for nonlinear fluorescence excitation through multiple scattering media," Opt. Lett. **43**(12), 2831 (2018).
40. H. B. de Aguiar, S. Gigan, and S. Brasselet, "Polarization recovery through scattering media," Sci. Adv. **3**(9), e1600743 (2017).

41. M. Kadobianskyi, I. N. Papadopoulos, T. Chaigne, R. Horstmeyer, and B. Judkewitz, "Scattering correlations of time-gated light," Optica **5**(4), 389 (2018).